# Understanding the wetting of transition metal dichalcogenides from an *ab initio* perspective

Siheng Li,[1] Keyang Liu,[2] Jiří Klimeš,[3] and Ji Chen[2,4,5,6,*]

[1]*International Center for Quantum Materials, School of Physics, Peking University, Beijing 100871, People's Republic of China*
[2]*School of Physics, Peking University, Beijing 100871, People's Republic of China*
[3]*Department of Chemical Physics and Optics, Faculty of Mathematics and Physics, Charles University, Prague 121 16, Czech Republic*
[4]*Collaborative Innovation Center of Quantum Matter, Beijing 100871, People's Republic of China*
[5]*Interdisciplinary Institute of Light-Element Quantum Materials and Research Center for Light-Element Advanced Materials,
Peking University, Beijing 100871, People's Republic of China*
[6]*Frontiers Science Center for Nano-Optoelectronics, Peking University, Beijing 100871, People's Republic of China*

Transition metal dichalcogenides (TMDs) are a class of two-dimensional (2D) materials that have been widely studied for emerging electronic properties. In this paper, we use computational simulations to examine the water adsorption on TMDs systematically and the wetting property of tungsten diselenide ) specifically. We start with density functional theory (DFT) based random phase approximation (RPA), assessing the performance of exchange-correlation functionals and comparing water adsorption on various TMDs. We also perform *ab initio* molecular dynamics (AIMD) simulations on WSe$_2$, from which we find that the distribution of interfacial water is sensitive to the exchange-correlation functional selected and a reasonable choice leads to a diffusive contact layer where water molecules prefer the "flat" configuration. Classical molecular dynamics (MD) simulations of water droplets on surfaces using appropriately parameterized water-surface interaction further confirm the dependence of water contact angle on the interaction and the interfacial water structure reproduced by different DFT functionals. Our study highlights the sensitivity of wetting to the water-substrate interaction and provides a starting point for a more accurate theoretical investigation of water-TMD interfaces.

DOI: 10.1103/PhysRevResearch.5.023018

## I. INTRODUCTION

Wetting of materials surface is one of the most ubiquitous phenomena in nature and has a pivotal influence on modern technologies [1]. Wetting of metal is widely known to promote corrosion and hamper lubrication [2]; wetting of clay and aerosol particles is fundamental to precipitation and ice nucleation [3]; wetting of oxide surfaces directly affects the photocatalysis of water and other molecules [4], to name just a few. Macroscopically, materials surfaces can be generally classified as hydrophobic or hydrophilic, providing the most basic understanding of wetting. In the past few decades, with the help of high-resolution surface probing techniques and computational simulations, the knowledge of wetting has been significantly deepened to the microscopic level [5].

Meanwhile, since the experimental discovery of graphene in 2004, the materials science community has witnessed an abrupt shift from bulk materials toward two-dimensional (2D) materials [6]. In the prospect of technological applications of two-dimensional materials, many other 2D materials have been investigated, among which 2D transition metal dichalcogenides (TMDs) have received a particularly large amount of interest because of the promising electronic behaviors as semiconductors, superconductors, and lasers, etc. [7,8]. At the same time, TMDs have also been investigated as molecular sensors, water purification membranes, and lubricants [9–11]. In these versatile applications of TMDs, wetting can play a very important role in their performance.

There are already a significant amount of efforts on establishing the macroscopic wettability of 2D TMDs, e.g., by measuring [12–15] and simulating [16,17] the contact angle of water droplets. However, it turns out as the description of wettability is very challenging because the macroscopic observations are very sensitive to the microscopic interaction between water and substrate. Taking the contact angle described by Young's equation as the model for wetting, a small change of water-surface interaction by tens of meV would change the substrate from super-hydrophilic to super-hydrophobic [16], which has also been revealed on graphene [18]. In recent experimental studies, force curve imaging of liquid water layer has been carried out on 2D materials such as graphene and TMDs, which further highlights the sensitivity of water-surface interaction to the interfacial water structure [19]. Moreover, water-TMD interactions are relatively weak, dominated collectively by weak charge transfers and long-range dispersion interactions, which are notoriously difficult to measure experimentally and calculate theoretically. From the theoretical perspective, there is a large variation of interaction energies between water and TMDs reported in the literature, ranging from 30 meV to 250 meV depending on the methods employed [20–22].

---

*ji.chen@pku.edu.cn

In this study, we carry out a systematic density functional theory study of water adsorption on TMDs. We start with the adsorption of water monomer on TMDs, where we compute the adsorption energy accurately using random phase approximation (RPA) as the benchmark to discuss the performance of exchange-correlation functionals of DFT for different TMDs. Then we adopt the best-performing functional to carry out a systematic study of water adsorption on TMDs. Following this, we carry out *ab initio* molecular dynamics (AIMD) simulations of liquid water-TMD interface with different functionals, and discuss the sensitivity of interfacial water structures to different methods, which is further explored by classical molecular dynamics (MD) with various parameters to reproduce the density profiles from AIMD.

## II. METHODS AND COMPUTATIONAL DETAILS

Geometry optimizations and adsorption energy calculations are carried out using the Vienna *ab initio* simulation package (VASP) [23]. Projected augmented wave potentials are used [24]. For the substrate, we use a $2\sqrt{3} \times 2\sqrt{3} \times 1$ super cell with 16 Å of vacuum. The lattice constants are presented in Table S1 of the Supplemental Material (SM) [25], which includes additional references [26–29]. For an isolated gas phase $H_2O$ molecule, the same box as that of the substrate is used. The computational approach used to obtain the benchmark value is combining the correlation energy from RPA [30–32] with the Hartree-Fock exchange energy plus the singles contribution (RSE) [33,34] based on PBE orbitals, i.e., the exact exchange energy EXX+RSE@PBE plus the correlation energy RPA@PBE. In this paper we dub this approach as "RPA". In our calculations, the adsorption energy is converged within several meV with a $2 \times 2 \times 1$ k-point mesh for the RPA correlation energy calculation as well as for the mean-field energy EXX+RSE. DFT geometry optimizations and energy calculations are performed with a $6 \times 6 \times 1$ k-point mesh without extra statements and an energy cut-off of 400 eV, which is determined by tests as shown in SM [25]. The exchange-correlation functionals considered include local density approximation (LDA), generalized gradient approximation (GGA), and meta-GGA [35,36]. Long-range dispersion interactions are included using various commonly adopted schemes (D2, D3, TS, rVV10, vdW-DF) [37–43].

AIMD simulations are carried out by VASP using the SCAN [36], rev-vdW-DF2, and PBE functionals. The simulations are performed for 80 ps and the last 50 ps were used for analyses for each functional. A timestep of 1 fs was used. The simulated system contains a $2\sqrt{3} \times 2\sqrt{3} \times 1$ slab of TMDs, 84 water molecules, and a 16-Å thick vacuum slab. The NVT ensemble was simulated via the Nosé-Hoover thermostat connected to all atoms in the system with a Nosé-mass corresponding to a period of 40 timesteps. The temperature for all AIMD simulations is 350 K. We note that in AIMD simulations of water, higher temperatures than 300 K are often used to reproduce the structure of water at room temperature because of the overestimated melting temperature by DFT functionals.

Classical MD simulations are performed using LAMMPS [44]. An NVT ensemble is used to simulate the droplet on the surface at 300 K. The Nosé-Hoover thermostat with a chain length of 10 and a relaxation time of 0.5 ps is used. Initial equilibration is performed for 125 ps with a timestep of 5 fs. The water-water interactions are described by a coarse-grained water model, namely the "mW" model [45], whereas the water-surface interaction is modeled by a Morse potential, $E = D_0[e^{-2\alpha(r-r_0)} - 2e^{-\alpha(r-r_0)}]$. For the system used to reproduce the water density profile for parameters identifications, we model liquid water with 756 water molecules on a 34.461 Å × 29.844 Å substrate. The simulations are performed for 10 ns with a time step of 10 fs for each functional. The droplets containing 11 613 water molecules were modeled on a 22.974 nm × 22.974 nm substrate. The initial droplet was generated with a water density of 1 kg/m$^3$. We then carried out simulations of droplets, which were initially placed above the substrate, for 20 ns with a time step of 10 fs. Tests were also performed to confirm that changing the shape and height of the initial droplets does not affect the shape of the droplet after equilibrium. The contact angles are estimated by simple geometric measurement.

## III. RESULTS

### A. Water monomer adsorption

The adsorption energy of water is the key quantity that indicates the hydrophobic/hydrophilic behavior of water, and it has been proven that the adsorption energy is very sensitive to the choice of exchange-correlation functional for water adsorption on graphene [18]. Therefore, before we discuss the adsorption of water clusters on $WSe_2$ and the aqueous interfaces, we first discuss stable water adsorption models, compare them for different TMDs substrates, and perform a benchmark examination for $WSe_2$ and $MoS_2$ as examples. To this end, we choose RPA as our benchmark method, which has been applied successfully to similar problems in the past decade, such as the adsorption of CO on metallic surfaces, the adsorption of water on graphene, and other bulk interactions [18,31,46,47]. RPA captures dynamic correlations explicitly, hence can lead to accurate calculation of dispersion interactions down to the sub-chemical accuracy level. Compared with other electronic structure methods, such as CCSD(T) and quantum Monte Carlo, RPA is computationally more affordable for the systems considered in this paper.

In general, the total adsorption energy is defined as follows:

$$E_{ad} = [E_{nH_2O@sub} - (nE_{H_2O} + E_{sub})]/n, \quad (1)$$

where $E_{nH_2O@sub}$ is the total energy of the substrate adsorbed with $n$ water molecules, and the energy of an isolated water molecule and the substrate correspond to $E_{H_2O}$ and $E_{sub}$, respectively. In this definition, negative adsorption energy is identified as a stable adsorption process, and larger/smaller absolute adsorption energy means over/underbinding. For water monomer adsorption, i.e., $n = 1$, the adsorption is fully contributed from the substrate. While for more water molecules, the adsorption energy generally includes the hydrogen bonds within the formed water cluster and the adsorption of the water cluster to the substrate.

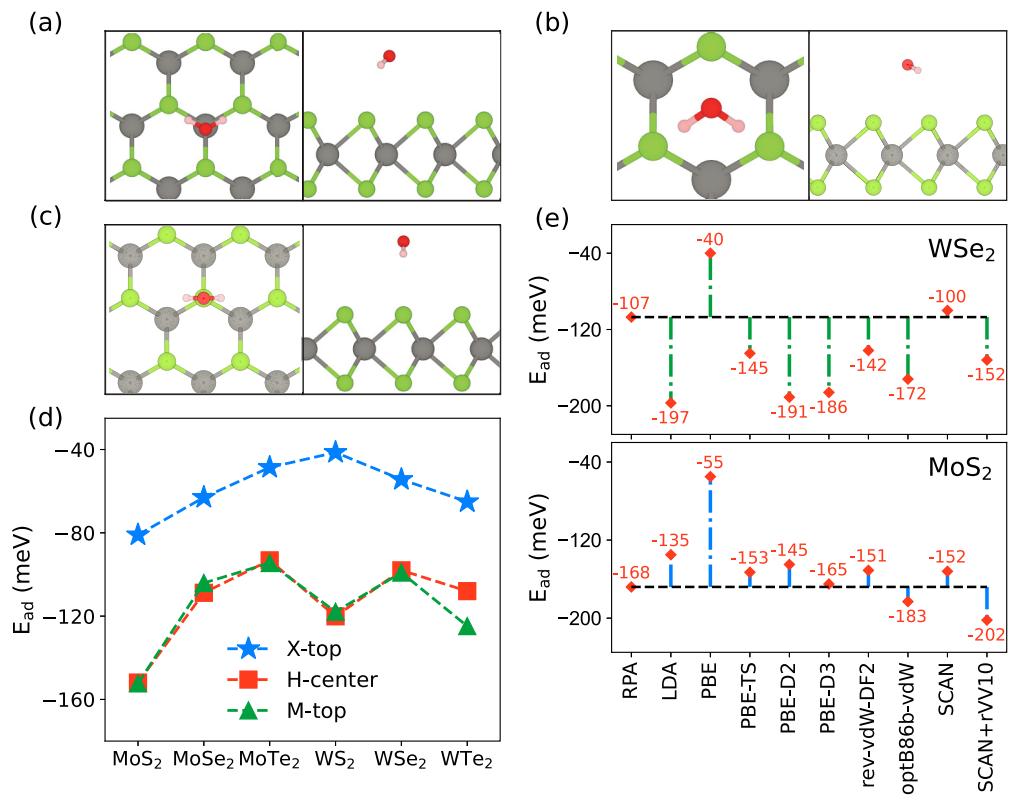

FIG. 1. Top and side views of stable adsorption configurations on the (a) W-top, (b) H-center, and (c) Se-top sites of WSe$_2$, where green, grey, pink, and red balls represent the selenium, tungsten, hydrogen, and oxygen atoms. (d) Adsorption energy of water monomer on the sites of X-top (blue star), H-center (red square), and M-top (green triangle) of MX$_2$ computed with SCAN. (e) Adsorption energy of water monomer on the H-center site of WSe$_2$ (upper panel with green lines) and MoS$_2$ (lower panel with blue lines) using different methods. The horizontal black dashed line located at –107 meV and –168 meV corresponds to the RPA results of WSe$_2$ and MoS$_2$ respectively.

Now we consider the adsorption of water monomer on WSe$_2$ as an example. The three typical configurations of water monomer adsorbed on WSe$_2$ are shown in Figs. 1(a)–1(c), which are optimized using the SCAN functional. These structures correspond to three different adsorption sites of the water molecule, namely the (a) W-top, (b) hexagon-center (H-center), and (c) Se-top. On WSe$_2$, the W-top, as well as the H-center configurations, are much more stable than the Se-top structure. Various other adsorption sites and water orientations were considered as the initial structures, and we did not observe more stable configurations for monomer adsorption.

Figure 1(d) presents a comparison of the monomer adsorption on six different TMDs computed using SCAN. We denote the TMDs substrates as MX$_2$, where M represents transition metals W or Mo and X is identified as chalcogen elements S, Se, or Te. Comparing the three adsorption sites, the H-center, and the M-top sites are always more stable than the X-top site by approximately 50–80 meV. Overall, we find that apart from MoS$_2$ the monomer adsorption on other TMDs is very similar, the interaction differs by less than 30 meV for the same adsorption site. It is worth noting that such a seemingly small difference, if translated to the hydrophobicity of material, might lead to a change of water contact angle by 20–30 degrees according to previous simulations [18], hence in such studies accurate calculation of water-substrate interaction is of great importance. Table S8 within SM [25] reports the distances of water molecules to the substrate on different adsorption sites, where we do not see activation of OH bonds and hence we expect the water-TMD interaction is dominated by weak hydrogen bonds and dispersion interactions.

In Fig. 1(e), we plot the adsorption energies of one water molecule on the H-center site of WSe$_2$ (upper panel with green stems) and MoS$_2$ (lower panel with blue stems) using RPA and a few commonly used DFT functionals, including LDA, PBE, PBE-TS, PBE-D2, PBE-D3, rev-vdW-DF2, optB86b-vdW, SCAN and SCAN+rVV10. Taking the RPA calculation with an adsorption energy of –107 meV for WSe$_2$ and –168 meV for MoS$_2$ as a reference, we can see which functional under/overestimates the adsorption energy. For WSe$_2$, the adsorption energies of LDA and PBE, i.e., –197 meV and –40 meV, represent the typical overbinding and underbinding cases, respectively. For MoS$_2$, interestingly, both LDA and PBE underestimate the adsorption, with stronger binding of 80 meV for LDA than for PBE. In Fig. S1 within SM [25] we show monomer absolute adsorption energy computed using various DFT functionals on all six substrates considered. We can see that the difference in performance is more related to the chalcogenide element, S, Se, and Te. This is because of the hydrogen-bond-like interactions between water and substrate, which is an important contribution to the total interaction. And the hydrogen-bond-like interaction depends on the electronegativity of the chalcogenide atom. Therefore, different DFT functionals can have quite differ-ent performances. These highlight the inaccuracy of simple semi-local DFT functionals, particularly the inconsistency of their performances. Including specific treatment on dispersion interaction improves the behavior of DFT functionals and yields better agreement with the reference. Nevertheless, these results further emphasize the importance of performing benchmark calculations beyond DFT when studying such problems.

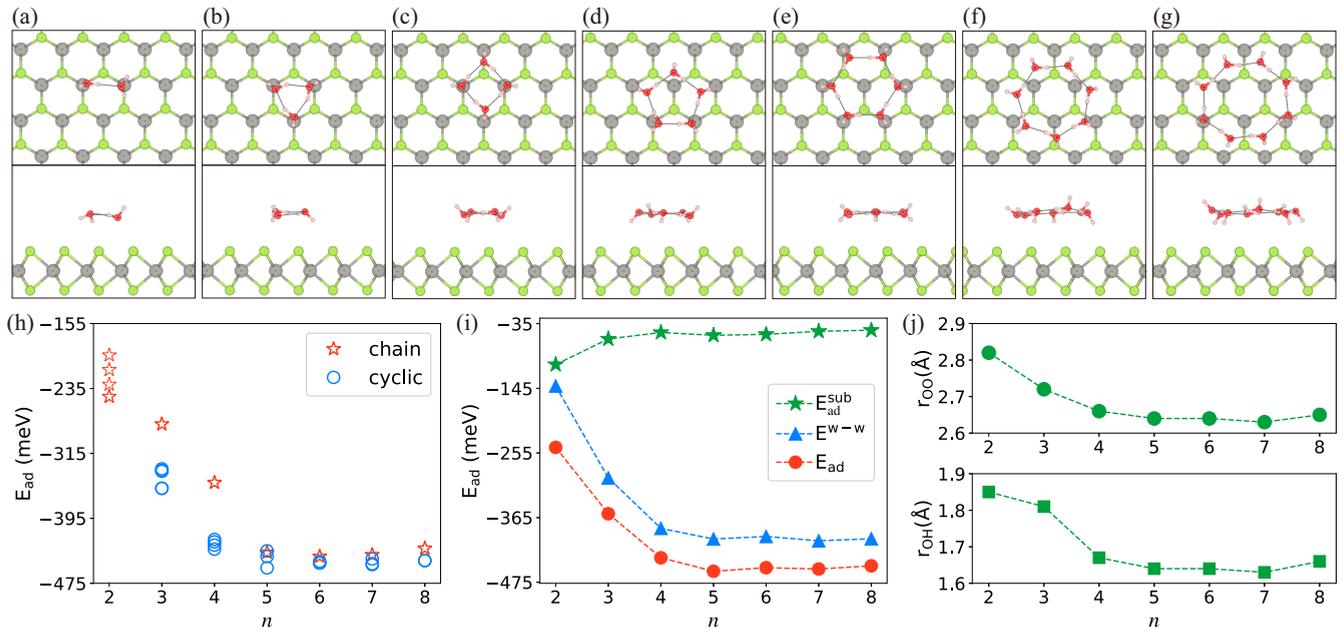

FIG. 2. [(a)–(g)] Top and side views of the lowest energy adsorbed $H_2O$ structures identified on $WSe_2$. (h) The adsorption energy as defined in Eq. (1) for 2–8 water molecules in several configurations, in which the red stars and blue circles represent the chain-like and cyclic structures [see Figs. S3– S9 within SM [25] for other structures not presented in (a)–(g)]. (i) The total and decomposed interaction energies as a function of the number of $H_2O$ (denoted as $n$). The total interaction, i.e., the adsorption energy is shown in the red curve, the water-substrate and water-water interactions are shown in the green and blue curves, respectively. (j) The averaged distance of oxygen atoms ($r_{OO}$) and hydrogen bonds ($r_{OH}$) between the nearest $H_2O$ as a function of $n$, as depicted with green circles and green square.

Among all the functionals tested, SCAN provides the best agreement with the RPA result only with an error within 10 meV for $WSe_2$. While for $MoS_2$, PBE-D3 performs the best only with an error within 5 meV and the error of SCAN is within 20 meV, which also performs as well as PBE-TS and rev-vdW-DF2. Consequently, including the long-range vdW interaction, as in SCAN-rVV10, would lead to larger deviations, since SCAN already captures largely the dispersion interactions for short separations. In the following sections, we mainly use SCAN to study the wetting behavior of water on $WSe_2$. Although SCAN still slightly underestimates, the comparison between different materials and different sites should be qualitatively reasonable. Note that SCAN is also a relatively good functional for water-water interaction, which only shows a minor over-structure feature for liquid water [48–50].

### B. Adsorption of water clusters

The adsorption of water monomer discussed above is the most fundamental microscopic aspect of wetting. However, the macroscopic phenomenon of wetting is related to the liquid-solid interface. From the microscopic point of view, the competition of the water-water interaction and the water-substrate interaction is the key to wetting. To provide insights into the understanding of the competition between water-water interaction and water-substrate interaction, we look at the formation of water clusters in this section. We use $WSe_2$ together with the SCAN functional as an example (see Table S9 within SM [25] for data calculated using rev-vdW-DF2 functional).

To identify the most stable water clusters adsorbed on $WSe_2$, we first built initial structures based on the most stable configuration of monomer adsorption, putting water molecules on the W-top, Se-top, and H-center sites, at the same time maximizing the possibility of forming hydrogen bonds. In addition, random arrangements of the structures are also applied to generate more initial structures. Then geometry optimization is performed to identify the most stable structures as presented in Figs. 2(a)–2(g) corresponding to 2–8 $H_2O$ molecules. The most stably adsorbed $H_2O$ dimer structure is displayed in Fig. 2(a). The adsorption energy per $H_2O$ as identified in Eq. (1) is −246 meV. In this structure both $H_2O$ molecules adsorb above the W-top sites. The $H_2O$ molecule that donates the hydrogen bond noticeably presents a so-called 0-leg structure, while the hydrogen acceptor water molecule has the 2-leg configuration. The height of the O atom in the donor and the acceptor from the substrate is 2.83 Å and 3.14 Å, respectively. The O–O distance in this configuration is 2.82 Å.

For larger water clusters, we consider 3–8 $H_2O$ molecules in chain-like and cyclic configurations. We find that the most stable structures are characterized as cyclic structures as shown in Figs. 2(b)–2(g), in which the amount of hy-drogen bonds is equal to those of H$_2$O molecules and each H$_2$O molecule donates and accepts one hydrogen bond. Water molecules prefer the W-top and H-center site to the Se-top site, so the optimized cyclic structures always remain as close as possible to W-top site for the small cluster. When the cluster is large, some water molecules have to move farther away from the W-top site to H-center site to maintain the cyclic structure. The adsorption energy per H$_2$O of all considered structures are plotted in Fig. 2(h). All the cyclic clusters have stronger interactions to WSe$_2$ than the chain-like structures. When it reaches five water molecules, the adsorption energy per H$_2$O reaches the lowest limit and only varies by a few meV when further increasing the number of H$_2$O. The tendency of the stablest configuration and the small variation of the adsorption energy from pentamer to octamer implies that the contact layer at the water-TMD interfaces may be quite flexible and does not feel the strong template effects of TMD substrates. In Table S9 within SM [25], we show calculations using the rev-vdW-DF2 functional. Similar to the monomer case the interaction energies are larger, but, nonetheless, we can reach the same conclusion that cyclic clusters from pentamer to octamer are the most stable ones with similar interaction energies. We note that similar trends have been reported for water cluster adsorption on the salt surface [51,52].

To further understand such behavior, we decompose the total interaction energy for each configuration into two parts, namely the hydrogen bond between water molecules and the interaction energy between water cluster and substrate, as plotted in Fig. 2(i). Specifically, we define

$$E^{\text{w-w}} = \left[E_{n\text{H}_2\text{O}} - nE_{\text{H}_2\text{O}}\right]/n \quad (2)$$

as the interaction among the water molecules and

$$E^{\text{sub}}_{\text{ad}} = \left[E_{n\text{H}_2\text{O@sub}} - (E_{n\text{H}_2\text{O}} + E_{\text{sub}})\right]/n \quad (3)$$

as the water-substrate interaction, where $E_{n\text{H}_2\text{O}}$ is the total energy of the $n\text{H}_2\text{O}$ cluster. The $n\text{H}_2\text{O}$ cluster has the same structure as its adsorption configuration, and the energy is obtained without further geometry optimization. From Fig. 2(i) we can see that the interactions contributed by hydrogen bonds are much stronger compared to the adsorption by the substrate. The weak adsorption by the substrate even decreases in magnitude with increasing $n$ and stays approximately constant for $n \geqslant 4$. On the other hand, the strong interaction energy of hydrogen bonds is strengthened significantly as the number of water molecules grows and reaches a plateau with $n$ increasing to 4. This implies that for a few water molecules on WSe$_2$ the weak adsorption by the substrate competes with the strong interaction of hydrogen bonds, whereas with increasing H$_2$O the hydrogen bonds dominate. Figure 2(j) demonstrates the averaged distances of the nearest oxygen atoms (denoted as $r_{\text{OO}}$) and the hydrogen bonds (denoted as $r_{\text{OH}}$) as a function of $n$. Compared to dimer and trimer, the clusters with four and more water molecules have shorter $r_{\text{OO}}$ and $r_{\text{OH}}$, whose differences are negligible with increasing $n$. Such results are consistent with the analysis that the interaction energy contributed by hydrogen bonds completely wins the competition and dominates for $n \geqslant 4$.

### C. Liquid water-TMD interface from AIMD

Now we move to AIMD simulations of liquid water on WSe$_2$, one that has the medium adsorption energy among the six TMDs investigated in Sec. III A. The water density probability distribution profiles $\rho(z)$, along the normal direction of the interface, are plotted in Fig. 3(b). The $\rho(z)$ profile from SCAN (blue line) features the contact layer of liquid water with a first peak at 3.6 Å away from the surface, which is consistent with the experimental measurements of the force-distance curve by Uhlig et al. [19]. The second peak of the $\rho(z)$ profile is at 6.7 Å and the third peak is around 10 Å, providing references for future experimental measurements, which are more challenging than the first layer because of the shift between the force-distance curve and the $\rho(z)$ profile. Interestingly, the first interfacial layer shows a broad peak with almost the same height as the second peak, indicating a diffusive contact layer on WSe$_2$. We have also performed simulations using different functionals, namely rev-vdW-DF2 (grey line) and PBE (pink line), to examine whether the diffusive contact layer is sensitive to the interaction energy between water and WSe$_2$ from different functionals. Despite rev-vdW-DF2 being the second best functional identified, it overestimates the interaction by about 35 meV, which seems to be large enough to induce a normal contact layer on WSe$_2$ surface as on other materials such as graphene or hBN [53]. As PBE further underestimates the interactions by over 60 meV, the interfacial water tends to diffuse away from the surface much further than SCAN does and features a broad indistinct contact layer.

To further explore the behavior of the contact layer, which reflects the most prominent impact of water-substrate interaction, we show the in-plane distributions of contact layer water molecules from the results of rev-vdW-DF2 and SCAN for comparison as shown in Fig. 3(c). The contact layer includes water molecules from the substrate to the first valley in the density profile in Fig. 3(b). The distribution from rev-vdW-DF2 presents an organized pattern that the water density is much higher around the W-top and H-center sites than the Se-top sites. This is in accordance with the analysis of water adsorption in Sec. III A that the adsorption of water monomer on W-top and H-center sites is much stronger than on Se-top. The small variation of adsorption energy as a function of the cluster size identified in Sec. III B may also contribute to the diffusive contact layer with the in-plane disorder. For SCAN the distribution is relatively more disordered than that for rev-vdW-DF2 and there is an unsharp preference for water molecules to stay near a certain atom. We attribute such a difference to the underestimated adsorption energy from SCAN compared to rev-vdW-DF2, which also leads to a diffusive contact layer with a broad peak instead of a regular contact layer with a sharper and higher peak.

The orientation of water molecules in the contact layer is further analyzed in Figs. 3(d) and 3(e). The angles are defined with respect to the surface normal, as depicted by the black arrow in the insets of Figs. 3(d) and 3(e), where $\theta$ represents the water dipole orientation (red arrow) and $\phi$ is the angle that measures the H-H direction (blue arrow). $\theta = 90°$ means the dipole of the water molecule is parallel to the surface and for zero degrees the dipole is perpendicular to the substrate.

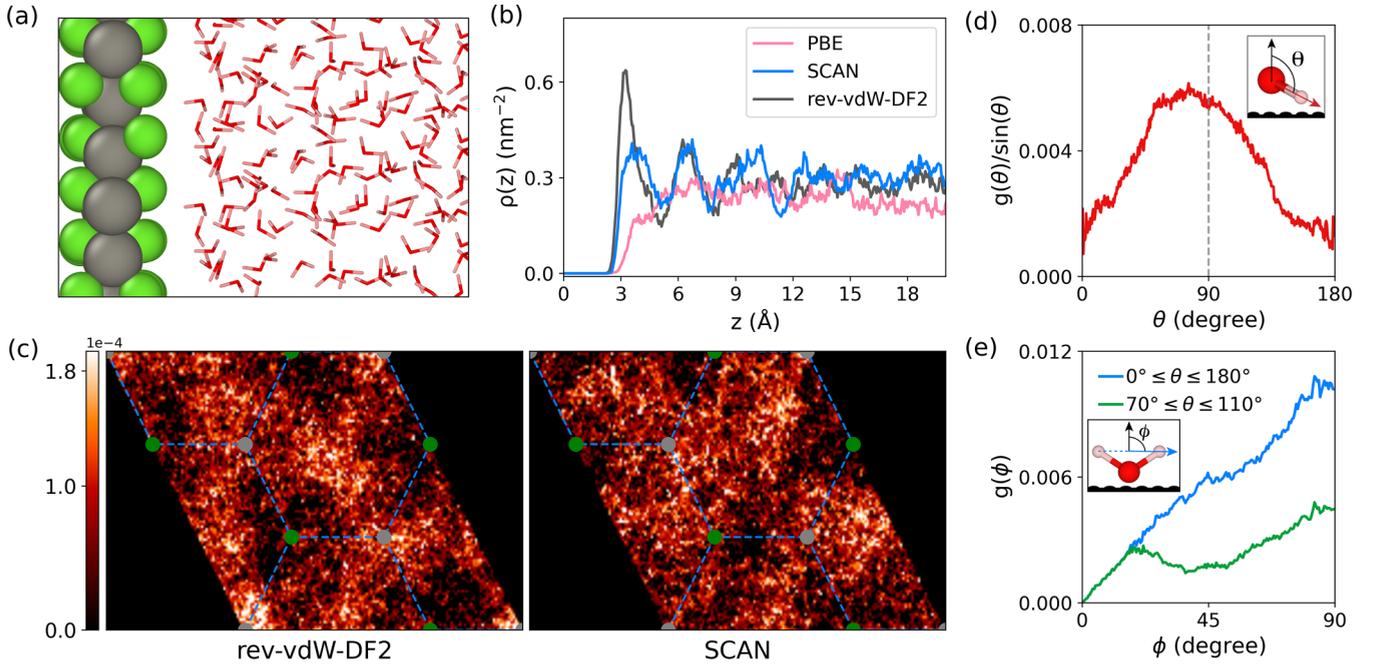

FIG. 3. AIMD simulations of liquid water on WSe$_2$. (a) A snapshot 2 from AIMD simulation of interfacial water obtained with OVITO [54]. (b) Water density profiles as a function of distance from the substrate as results of PBE, SCAN and rev-vdW-DF2. The zero of height refers to the average height of the top planar of Se atoms. (c) Water distributions of the contact layer within $\sqrt{3} \times \sqrt{3} \times 1$ hexagonal supercell from rev-vdW-DF2 (left) and SCAN (right). The grey and green circles show the averaged position of surface W and Se atoms respectively. Angle distribution profiles of (d) dipole orientation $\theta$ and (e) leg rotation $\phi$ for the contact layer water. The blue and green curves in (e) correspond to the water within contact layer with water orientation of $0° \leqslant \theta \leqslant 180°$ (all) and $70° \leqslant \theta \leqslant 110°$ (flat orientation) respectively. The insets of (d) and (e) are the illustrations of dipole orientation (red arrow) and leg rotation H-H direction (blue arrow) with respect to the surface normal (black arrow) respectively.

Here we consider water with $70° \leqslant \theta \leqslant 110°$ as "flat" dipole. The maximum of $\phi_{max} = 90°$ is that two H atoms are at the same distance away from the surface and deviation from $90°$ means rotation along the dipole axis. We note that the minimum angle of H-H direction $\phi_{min}$ is restricted by $\theta$ with a relation of $\phi_{min} = |\theta - 90°|$, meaning that H-H direction can only be parallel to the surface for water molecule with the perpendicular dipole. The two distributions together determine the water orientation. In the contact water layer, the most favorable orientation is $50° \leqslant \theta \leqslant 100°$ for the dipoles and around $90°$ for the H-H direction as plotted with red and blue lines in Figs. 3(d) and 3(e) respectively. $\theta$ does not strictly maximize at $90°$ means that the water molecules have their dipoles slightly tilted away from the surface, which is different from the adsorption configuration of a single water molecule. In particular, the favorite H-H direction for "flat" dipoles as defined above is around $90°$ as plotted with the green line in Fig. 3(e), which is denoted as a typical "flat" configuration or in-plane dipole.

Overall, we find the contact layers at the WSe$_2$ interface consists of water molecules that prefer in-plane dipole orientations. Such orientation feature is qualitatively similar to other hydrophobic surfaces such as graphene/hBN [55], and shall lead to a significant reduction of interfacial dielectric response. In this regard, there are two factors that come into play and may compete with each other. On the one hand, the water contact layer at WSe$_2$ interface is diffusive with a broad distribution perpendicular to the interface, meaning that water molecules can reorient themselves more easily. On the other hand, the water molecules prefer a "flat" configuration in the contact layer, which reduces the dielectric response. We also expect such analyses to be extended to other TMDs examined in this paper, which have similar interactions with water. Therefore, it will be interesting to quantitatively establish whether interfacial water on TMDs has larger or smaller dielectric response in the future.

### D. Modeling water droplet on substrate

In above sections we have focused on the discussion of water adsorption and AIMD simulations in periodic box, which provide information of the interfacial structure but do not directly connect to the wetting property of the substrate. This section aims at further establishing the impact of water-surface interaction on the contact angle of water droplet, which is often used as a measure of the hydrophobicity/hydrophilicity of material. As it is impossible to perform large water droplet simulations using DFT directly, we perform classical MD simulations of water droplet by varying the water-surface interaction parameters, including, but not limited to, the adsorption energy. Our strategy is that we first identify the set of parameters that can reproduce the interfacial water structure, then use the parameters to simulate the droplet and measure the contact angle.

The water-surface interaction is described by a Morse potential that reads $E = D_0[e^{-2\alpha(r-r_0)} - 2e^{-\alpha(r-r_0)}]$, where the energy depth parameter $D_0$ can be interpreted as the effective interaction of water to the substrate. Such simulations have been performed in a previous study of graphene [18], where $D_0$ was tuned according to the adsorption energy of water monomer. Here, we test various sets of Morse potential parameters aimed at reproducing the water density profile from AIMD simulations. Morse potential parameters found in such a manner can model the liquid water-TMD interface more accurately because water molecules at the liquid-TMD interface are different from the singly adsorbed water molecule on TMDs. Figure 4(a) illustrates the water density profiles from the MD simulations and the corresponding AIMD simulations using PBE, SCAN, and rev-vdW-DF2. The three settings are labeled with parameter $D_0$, namely "40 meV", "146 meV", and "340 meV". This indeed shows that water-TMD interaction in a liquid state is different from the adsorption of a single molecule. Figure 4(b) shows the snapshots obtained after the water droplet is stabilized in the simulation, in which 30°, 60°, and 109° correspond to the density profiles fitted to those of PBE, SCAN, and rev-vdW-DF2 from AIMD. We see that the three settings lead to completely different contact angles of water droplets. The simulation indicates an intriguing consequence that a difference of tens of meV for the monomer adsorption from different functionals in DFT might cause a great deviation of wetting behavior.

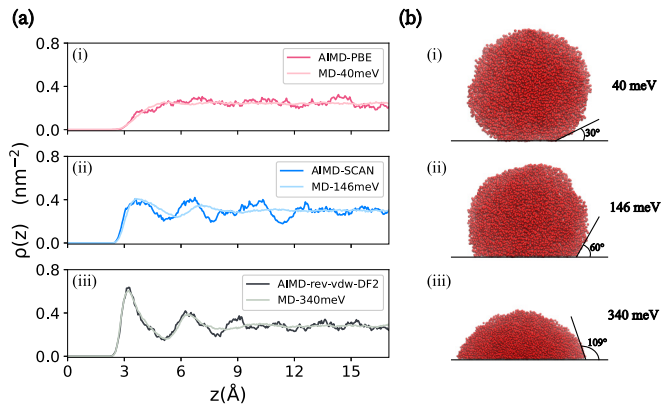

FIG. 4. MD simulations. (a) Water density profiles obtained with MD and AIMD. (b) Snapshots of liquid droplets from MD simulations, corresponding to the same water density profiles of panel (a). The three settings of MD simulations include the following parameters, (i) $D_0 = 40$ meV; $\alpha = 0.8$ Å$^{-1}$; $r_0 = 4.00$ Å, and (ii) $D_0 = 146$ meV; $\alpha = 0.8$ Å$^{-1}$; $r_0 = 3.59$ Å, and (iii) $D_0 = 340$ meV; $\alpha = 0.8$ Å$^{-1}$; $r_0 = 3.28$ Å.

## IV. CONCLUSIONS AND DISCUSSION

We have carried out a thorough investigation of water-TMD interfaces, and improved our understanding of the wetting behavior of a family of TMDs from *ab initio*. Our accurate adsorption energy calculation is based on RPA, a well-tested approach for such systems, which leads to the selection of the SCAN and rev-vdW-DF2 functionals for further examination of the molecular wetting process of $WSe_2$ and shows a different performance for $MoS_2$ and $WSe_2$. We find the most stable water cluster on $WSe_2$ is a cyclic pentamer, followed by larger clusters with very minor energy differences. AIMD simulations at water-$WSe_2$ interface by SCAN functional identify a diffusive contact layer as a direct consequence of the unique interaction between water and $WSe_2$. Following this, the MD simulations further confirm that different interaction energies may not only result in different density distributions in the contact layer but also lead to totally different contact angles of the water droplet.

Our study is a step further for developing water-TMD force fields using, e.g., machine learning methods, which will help to understand liquid-TMD slip, adhesion, and dielectric properties of interfacial water on TMDs, etc. However, it is worth noting that to achieve an accurate simulation of the water-TMD interface fully from *ab initio* is still challenging, which requires accurate treatment of both water-substrate interaction and water-water interaction. This study only focuses on the water-substrate interaction, while the performance of DFT functionals on liquid water has been discussed in the bulk of the literature. To conclude, the main aim of this paper is to highlight the importance of treating water-surface interaction accurately, so that in the future it can be considered on the same footing when discussing the modeling of water-TMD interfaces.

## ACKNOWLEDGMENTS

This work was supported by the National Key R&D Program of China under Grant No. 2021YFA1400500, the National Natural Science Foundation of China under Grants No. 11974024 and No. 92165101, and the Strategic Priority Research Program of Chinese Academy of Sciences under Grant No. XDB33000000. We are grateful for computational resources supported by High-performance Computing Platform of Peking University, the TianHe-1A supercomputer and Shanghai Supercomputer Center. J.K. was supported by the European Union's Horizon 2020 research and innovation programme via the ERC Grant APES (No. 759721).